\documentclass[journal=nalefd,manuscript=letter,layout=twocolumn]{achemso}

\usepackage{chemformula} % Formula subscripts using \ch{}
\usepackage[T1]{fontenc} % Use modern font encodings
\usepackage{amssymb}
\usepackage{bm}
\usepackage[breaklinks=true,colorlinks=true,linkcolor=blue,urlcolor=blue,citecolor=blue]{hyperref}

\newcommand*{\tempop}[3][\textstyle]{\settowidth{\dimen1}{$#1\hat{#2}$}\makebox[\dimen1][l]{$#1\hat{#2\mspace{#3}}$}}
\newcommand*{\xop}[1]{{\mathchoice{\tempop[\displaystyle]{#1}{3.5mu}}{\tempop{#1}{3.5mu}}{\tempop[\scriptstyle]{#1}{3.5mu}}{\tempop[\scriptscriptstyle]{#1}{3mu}}}}
\newcommand*{\chat}[1]{\ensuremath{\xop{#1}}}

\author{Jan-Philip Joost}
\affiliation{Institut f\"ur Theoretische Physik und Astrophysik, 
Christian-Albrechts-Universit\"{a}t zu Kiel, D-24098 Kiel, Germany}
\author{Antti-Pekka Jauho}
\affiliation{CNG, DTU Physics, Technical University of Denmark, Kongens Lyngby, DK 2800, Denmark}
\author{Michael Bonitz}
\affiliation{Institut f\"ur Theoretische Physik und Astrophysik, 
Christian-Albrechts-Universit\"{a}t zu Kiel, D-24098 Kiel, Germany}
\email{bonitz@theo-physik.uni-kiel.de}

\title{
Correlated Topological States in Graphene Nanoribbon Heterostructures
}% 

\begin{document}

%%%%%%%%%%%%%%%%%%%%%%%%%%%%%%%%%%%%%%%%%%%%%%%%%%%%%%%%%%%%%%%%%%%%%
%% The "tocentry" environment can be used to create an entry for the
%% graphical table of contents. It is given here as some journals
%% require that it is printed as part of the abstract page. It will
%% be automatically moved as appropriate.
%%%%%%%%%%%%%%%%%%%%%%%%%%%%%%%%%%%%%%%%%%%%%%%%%%%%%%%%%%%%%%%%%%%%%
\begin{tocentry}

\includegraphics[width=\textwidth]{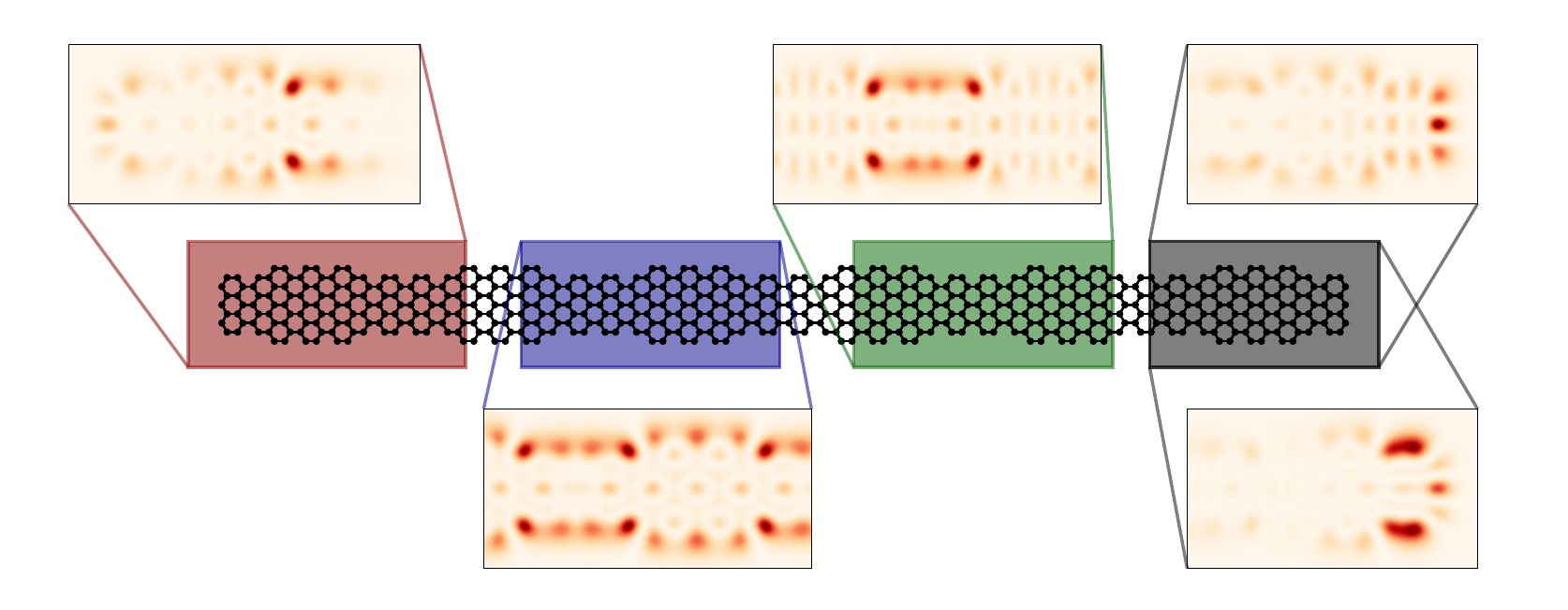}

\end{tocentry}

%%%%%%%%%%%%%%%%%%%%%%%%%%%%%%%%%%%%%%%%%%%%%%%%%%%%%%%%%%%%%%%%%%%%%
%% The abstract environment will automatically gobble the contents
%% if an abstract is not used by the target journal.
%%%%%%%%%%%%%%%%%%%%%%%%%%%%%%%%%%%%%%%%%%%%%%%%%%%%%%%%%%%%%%%%%%%%%
\begin{abstract}
Finite graphene nanoribbon (GNR) heterostructures host intriguing topological in-gap states  (Rizzo, D. J. et al.~
%Rizzo, D. J.; Veber, G.; Cao, T. et al. 
\textit{Nature} \textbf{2018}, \textit{560}, 204]). These states may be localized either at the bulk edges, or at the ends of the structure. Here we show that correlation effects (not included in previous density functional simulations)  play a key role in these systems: they  result in increased magnetic moments at the ribbon edges accompanied by a significant energy renormalization of the topological end states -- even in the presence of a metallic substrate.  Our computed results are in  excellent agreement with the experiments. Furthermore, we discover a striking, novel mechanism that causes an energy splitting of the non-zero-energy topological end states for a weakly screened system. 
%. We predict that states with similar features are present in other GNR heterostructures as well.
We predict that similar effects should be observable in other GNR heterostructures as well.
%Including quasiparticle corrections within the $GW$ approximation we observe
%[Rizzo, D. J.; Veber, G.; Cao, T.; Bronner, C.; Chen, T.; Zhao, F.; Rodriguez, H.; Louie, S. G.; Crommie, M. F.; Fischer, F. R. \textit{Nature} \textbf{2018}, 560, 204]
\end{abstract}

\begin{figure*}[t]
\includegraphics[width=\textwidth]{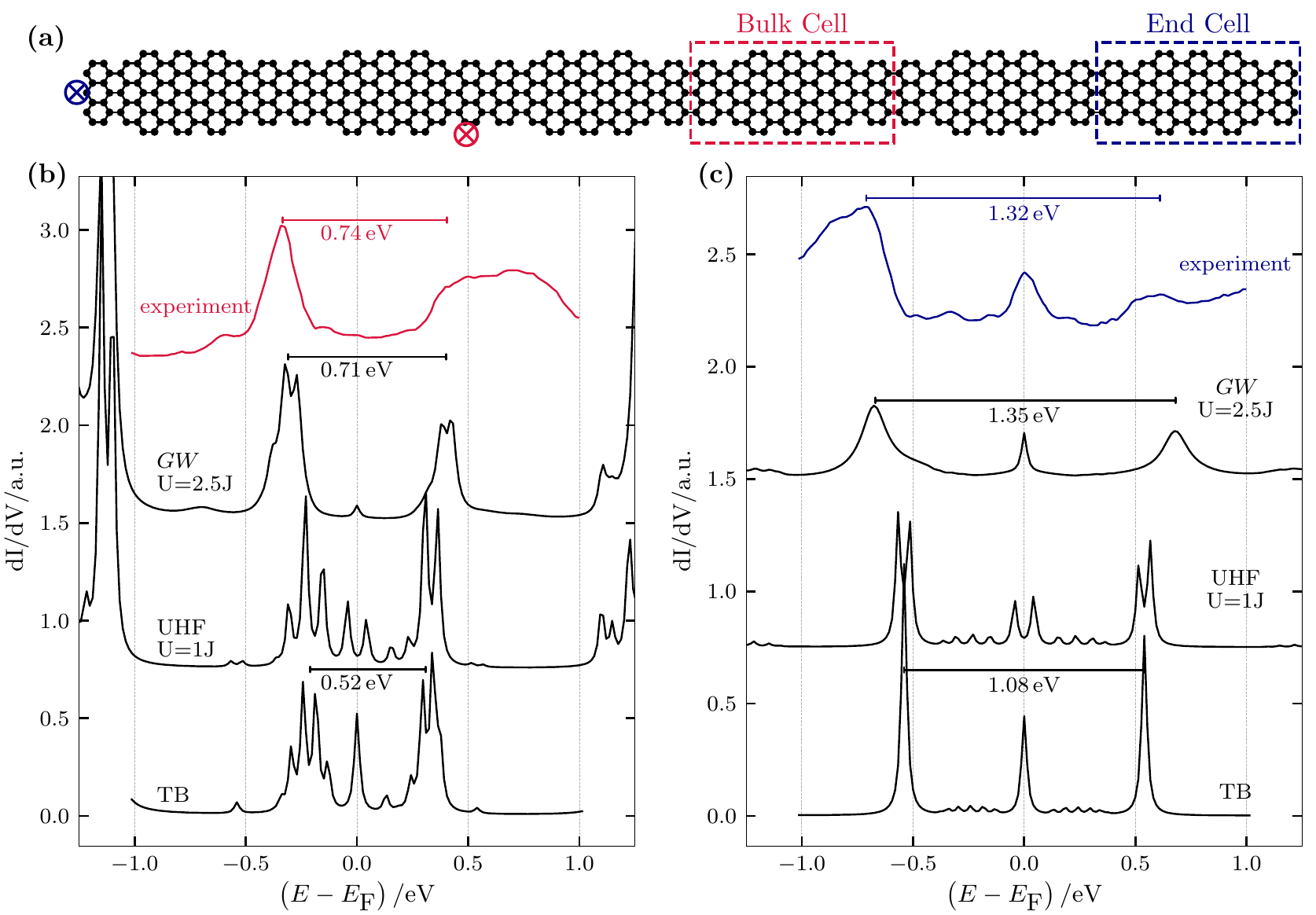}
\caption{\label{fig:exp_comparison} (a) 7-9-AGNR heterostructure containing six unit cells. The red [blue] cross marks the position of the d$I$/d$V$ spectra shown in (b) [(c)]. The red (blue) dashed rectangle marks the bulk (end) unit cell referenced in Fig.~\ref{fig:size}. (b) d$I$/d$V$ spectrum measured (red) and simulated (black) at the position in the bulk region marked by the red cross in (a). (c) d$I$/d$V$ spectrum measured (blue) and simulated (black) at the position in the end region marked by the blue cross in (a). The curves in (b) and (c) are shifted vertically, for better comparison. The experimental data are taken from Ref.~\citenum{rizzo_topological_2018} and corrected for charge doping effects.}
\end{figure*}

%%%%%%%%%%%%%%%%%%%%%%%%%%%%%%%%%%%%%%%%%%%%%%%%%%%%%%%%%%%%%%%%%%%%%
%% Start the main part of the manuscript here.
%%%%%%%%%%%%%%%%%%%%%%%%%%%%%%%%%%%%%%%%%%%%%%%%%%%%%%%%%%%%%%%%%%%%%
%\section{Introduction}
{\bf Introduction}.
Strongly correlated materials host exciting physics such as superconductivity or the fractional quantum Hall effect~\cite{stormer_fractional_1999}.
While in monolayer graphene electron correlations are weak~\cite{das_sarma_electronic_2011}, 
%due to the linear dispersion at low %energies
carbon based finite systems and heterostructures can exhibit flat bands near the Fermi energy resulting in nontrivial correlated phases~ \cite{neilson_many_body_2016,balzer_prl_18}.
One example are magic-angle twisted graphene bilayers where localized electrons lead to Mott-like insulator states and unconventional superconductivity~\cite{cao_correlated_2018,cao_unconventional_2018,kerelsky_maximized_2019}.
Another exciting class of systems that has been predicted to host strongly localized phases are graphene nanoribbon (GNR) heterostructures~\cite{cai_graphene_2014,chen_molecular_2015,cao_topological_2017,jacobse_electronic_2017,ma_seamless_2017,wang_quantum_2017,rizzo_length-dependent_2019}. Similar to topological insulators they combine an insulating bulk with robust in-gap boundary states~\cite{hasan_colloquium:_2010,asboth_short_2016} and are expected to host Majorana fermions in close proximity to a superconductor~\cite{groning_engineering_2018}. Recently, it was confirmed that GNR heterostructures composed of alternating segments of 7- and 9-armchair GNRs (AGNRs), as sketched in Fig.~\ref{fig:exp_comparison}(a), exhibit new topological bulk bands and end states that differ qualitatively from the band structures of pristine 7- and 9-AGNRs~\cite{rizzo_topological_2018}. 

Although electronic correlations are expected to play a crucial role for the localized topological states in 
%these systems
GNR heterostructures~\cite{lado_emergent_2019}, so far, most theoretical 
%modelling of GNR heterostructures 
work 
has been restricted to tight-binding (TB) models or density functional theory within the local density approximation (LDA-DFT) which are known to completely ignore or underestimate these effects. 
%Going beyond those simple theories, in this paper, we present simulations of 
Here, we present a systematic analysis of electronic correlations in 
7-9-AGNRs, based on a Green functions method with $GW$ self-energy~\cite{robert_book,Schluenzen2019} applied to an effective Hubbard model.
We compute the differential conductance and 
%compare 
%our simulated dI/dV data 
%to 
find excellent agreement with
the experimental measurements of Ref.~\citenum{rizzo_topological_2018}. Our calculations reveal that, even in the presence of a screening Au(111) surface, local electronic correlations induce a strong energy renormalization of the band structure. Especially the topological end states localized at the heterostructure-vacuum boundary experience strong quasiparticle corrections which are not captured by LDA-DFT. For freestanding systems, or systems on an insulating surface, we predict that these states exhibit an energy splitting due to a magnetic instability at the Fermi energy. The local build-up of electronic correlations is further analyzed by considering the local magnetic moment at the ribbon edges. We also examine finite size effects and the origin of the topological end states by varying the system size and end configuration, respectively. On this basis we predict that a whole class of systems exists that can host end states with similar exciting properties. 
%
%Our calculations reveal that, even in the presence of a screening Au(111) surface, local electronic correlations induce a strong energy renormalization of the topological end states localized at the heterostructure-vacuum boundary which is not captured by LDA-DFT. This is accompanied by local magnetic polarization on all ribbon edges. For freestanding systems, or systems on an insulating surface, we predict that all three \textcolor{purple}{{\bf it is not clear what the three states are!}}  topological end states exhibit an energy splitting due to the magnetic instability of the zero-energy peak. Thus, the inclusion of electronic correlations allows us to achieve a detailed understanding of the mechanisms that lead to the creation of the end states. On this basis we predict that a whole class of systems exists that can host end states with similar exciting properties. 
%\textbf{this would be a good closing of the introduction. The following paragraph could be included in the previous one describing the results.}\\
%Our paper is organized as follows. We first present the system and the model Hamiltonian. Then, we describe the theoretical approach including the choice of parameters. Our results are first compared to the experimental measurements of Ref.~\cite{rizzo_topological_2018}. Thereafter, the role of electronic correlations and the system size dependence are discussed. Finally, we investigate the origin of the end states by considering different end configurations.\\

\textbf{Model.}
%The system we consider in this work is depicted in Fig.~\ref{fig:exp_comparison}a. The 7-9-AGNR heterostructure consisting of alternating 7-AGNR and 9-AGNR segments was realized experimentally by Rizzo et al.~\cite{rizzo_topological_2018}.
We consider the 7-9-AGNR heterostructure consisting of alternating 7-AGNR and 9-AGNR segments as depicted in Fig.~\ref{fig:exp_comparison}(a). The system was realized experimentally on a metallic Au(111) surface by Rizzo \textit{et al.}~\cite{rizzo_topological_2018} %They found that this particular heterostructure 
%exhibits 
who observed topological in-gap states 
%that emerge 
at the heterojunction between 7- and 9-AGNR segments (bulk), and at the termini of the heterostructure (end). While previous LDA-DFT simulations describe reasonably well the bulk bands, they do not reproduce quantitatively the experimental energies of the end states (see below). %Thus, for a more accurate treatment of the topological states an approach that includes electronic correlations beyond the level of LDA-DFT is needed. 
%To this end, 
To overcome these limitations, we apply a recently developed Green functions approach which gives access to spectral and magnetic properties of the system, see,~ e.g.~Refs.~\citenum{haug_2008_quantum,balzer-book,robert_book}. The electronic system is described with an effective Hubbard model, the Hamiltonian of which is expressed in terms of the  operators ${\chat{c}}_{\bm{i}\alpha}^\dagger$ and ${\chat{c}}_{\bm{j}\alpha}$ that create and annihilate an electron with spin projection $\alpha$ at site $i$ and $j$, respectively,
\begin{equation}
\chat{H} = -J\sum_{\langle\bm{i},\bm{j}\rangle,\alpha} {\chat{c}}_{\bm{i}\alpha}^\dagger{\chat{c}}_{\bm{j}\alpha}+U\sum_{\bm{i}}{\chat{c}}_{\bm{i}\uparrow}^\dagger{\chat{c}}_{\bm{i}\uparrow}{\chat{c}}_{\bm{i}\downarrow}^\dagger {\chat{c}}_{\bm{i}\downarrow}\,,
\label{eq:hubbard}
\end{equation}
where $J=2.7\,\mathrm{eV}$ is the hopping amplitude between adjacent lattice sites~\cite{reich_tight_binding_2002}, and $U$ is the on-site interaction. The edges of the GNR are assumed to be H-passivated. 
%The model is solved using the Green function approach which gives access to spectral and magnetic properties of the system~\cite{}. 
Observables can be computed from the Green function $\bm{G}(\omega)$ that is defined in terms of the operators ${\chat{c}}_{\bm{i}\alpha}^\dagger$ and ${\chat{c}}_{\bm{j}\alpha}$, for details see the Supporting Information (SI).
Correlation effects are included via the self-energy $\bm{\Sigma}$ which enters in the self-consistent Dyson equation \cite{haug_2008_quantum, balzer-book}
\begin{equation}
 \bm{G}(\omega) = \bm{G}_0(\omega) + \bm{G}_0(\omega) \bm{\Sigma}(\omega) \bm{G}(\omega)\,,
 \label{eq:dyson}
\end{equation}
where the single-particle Green function $\bm{G}$ contains the spectral and magnetic information of the system, and $\bm{G}
_0$ is its non-interacting limit. In the present work we report the first full GW-simulations of the system (\ref{eq:hubbard}, \ref{eq:dyson}) for experimentally realized GNR heterostructures, as the one shown in  Fig.~\ref{fig:exp_comparison}(a). The details of the numerical procedure  are provided in  the SI.
%The solution procedure of Eq.~(\ref{eq:dyson}) is provided in  the SM.
\\

In what follows, we compare the tight-binding (TB), unrestricted Hartree-Fock (UHF), and $GW$ approximations for $\Sigma$, in order to quantify electronic correlation effects. The TB approach, corresponding to setting $U=0$, is often used to describe GNRs, due to its simplicity~\cite{xu_elementary_2007,sevincli_superlattice_2008,tao_spatially_2011,groning_engineering_2018}, and here serves as a point of reference for the uncorrelated system. 
% The on-site interaction for UHF was chosen as $U=1J$ which describes correctly the edge states in ZGNRs~\cite{yazyev_theory_2011,magda_room-temperature_2014}. \textcolor{purple}{the UHF approximation which is known to qualitatively describe edge magnetism in free-standing ZGNRs~\cite{magda_room-temperature_2014,van_der_lit_suppression_2013,wang_giant_2016}.}
For $GW$ the on-site interaction was %determined to 
chosen such that it
reproduces the experimental bulk band gap of Ref.~\citenum{rizzo_topological_2018}, resulting in $U=2.5J$, see SI. This choice of $U$ also takes into account screening effects of the metallic substrate. The description of free-standing GNRs within $GW$ requires a larger on-site interaction~\cite{Joost2019} which makes the selfconsistent solution of Eq.~(\ref{eq:dyson}) more challenging.  Nonetheless, to get a qualitative understanding of the properties of free-standing heterojunctions we employ the UHF approximation which is known to qualitatively describe edge magnetism in free-standing ZGNRs. For this case the on-site interaction was chosen as $U=1J$~\cite{yazyev_theory_2011}.
The spatially resolved d$I$/d$V$ data, recorded in an STM experiment,  are generated by placing 2$p_z$ orbitals on top of the atomic sites of the lattice structure, following the procedures described in Refs.~\citenum{tersoff_theory_1985,meunier_tight-binding_1998}.
 
% \begin{enumerate}[(1)]
% \setcounter{enumi}{-1}
%  \item Diagonalize single-particle Hamiltonian $\hat{h}$\\ and initialize $\hat{G}^{\mathrm{R}/\mathrm{A}}(\omega) = \hat{G}_0^{\mathrm{R}/\mathrm{A}}(\omega)$
%  \item Calculate $\hat{G}^{\gtrless}(\omega)$ from $\hat{G}^{\mathrm{R}/\mathrm{A}}(\omega)$
%  \item Perform FFT for $\hat{G}^{\gtrless}(\omega)$: $\omega \rightarrow t$
%  \item Calculate $\hat{\Sigma}^{\gtrless}(t)$ and $\hat{\Sigma}^{\mathrm{R}/\mathrm{A}}(t)$
%  \item Perform FFT for $\hat{\Sigma}^{\mathrm{R}/\mathrm{A}}(t)$: $t \rightarrow \omega$
%  \item Solve Dyson equation for $\hat{G}^{\mathrm{R}/\mathrm{A}}(\omega)$
%  \item Repeat steps (1)-(5) until convergence is reached
% \end{enumerate}

\begin{figure*}[t]
\includegraphics[width=\textwidth]{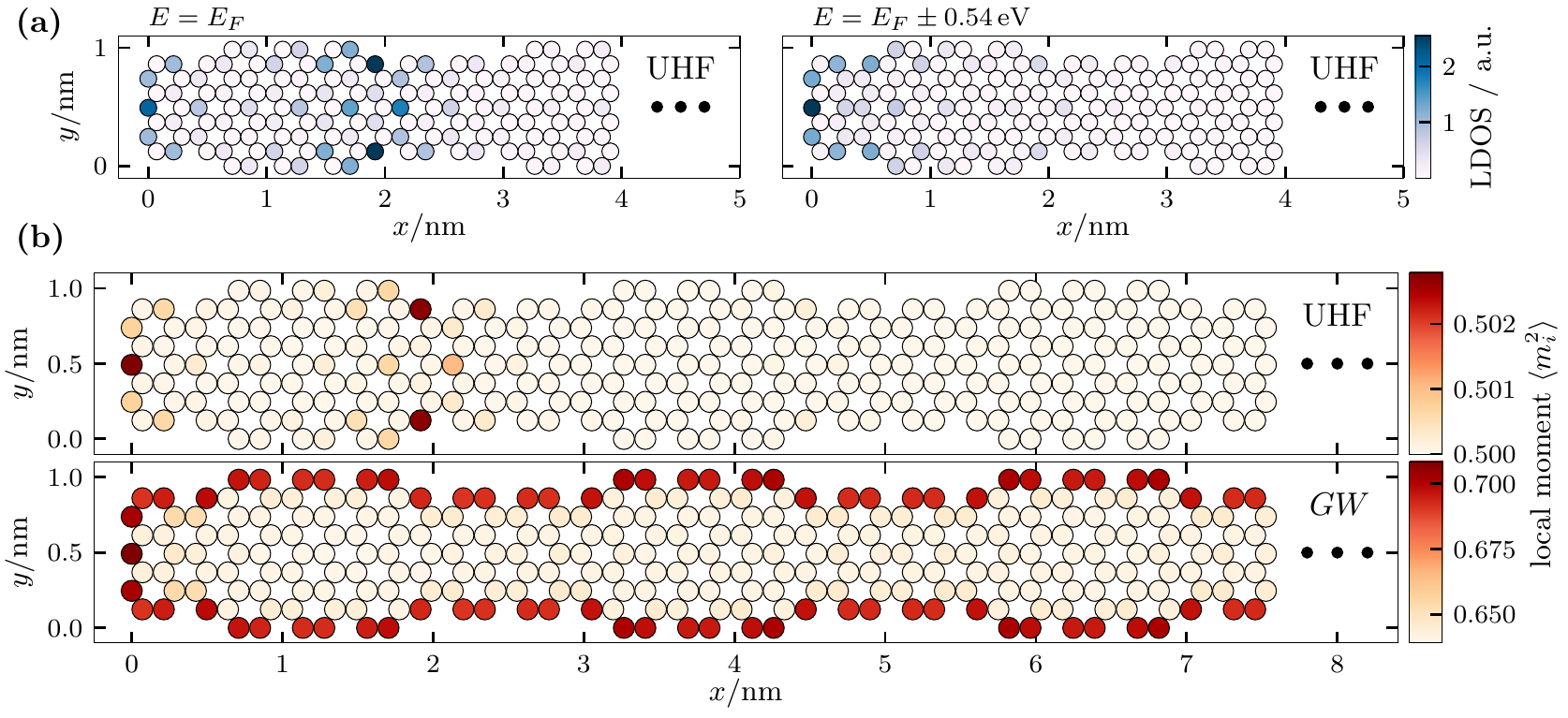}
\caption{\label{fig:localmoment} (a) LDOS of the topological end states at $E=E_F$ (left) and $E=E_F\pm0.54\,\mathrm{eV}$ (right) for UHF as shown in Fig.~\ref{fig:exp_comparison}(c). (b) Local moment $\langle m_i^2\rangle$, Eq.~(\ref{eq:moment}), of the 7-9-AGNR of Fig.~\ref{fig:exp_comparison}.(a)
%containing six unit cells 
calculated within UHF (top) and $GW$ (bottom). Since the ribbon is symmetric only three of the six unit cells are shown as indicated by the three black dots.}
\end{figure*}

\textbf{Quasiparticle renormalization}.
%To assess the influence of electronic correlations on the topological bulk and end states in the 
%7-9-AGNR heterostructure on Au(111), we compare in Figs.~\ref{fig:exp_comparison}b and c the results of 
%our approach using the 
%TB, UHF and $GW$ simulations with the local $dI/dV$ measurements of 
In Figs.~\ref{fig:exp_comparison}(b) and (c) we present differential conductance results for the 7-9-AGNR heterostructure on Au(111), comparing TB, UHF and $GW$ simulations to the experiment of 
Ref.~\citenum{rizzo_topological_2018}.\footnote{In the experiment the ribbon was slightly doped which shifts the d$I$/d$V$ spectrum to higher energies. Our calculations were performed for half filling. For comparison the experimental data was shifted so that the zero-energy peaks of theory and experiment match.} Our calculations were performed for a system containing six unit cells as shown in Fig.~\ref{fig:exp_comparison}(a). In accordance with the measurements the results for the bulk (end) calculation are averaged over the area marked by the red (blue) cross. In the experiment, a band gap of $E_\mathrm{g,bulk}^\mathrm{exp}=0.74\,\mathrm{eV}$ between the bulk bands is observed whereas the gap between the non-zero energy end peaks is $E_\mathrm{g,end}^\mathrm{exp}=1.32\,\mathrm{eV}$. The TB approximation vastly underestimates the gaps, with $E_\mathrm{g,bulk}^\mathrm{TB}=0.52\,\mathrm{eV}$ and $E_\mathrm{g,end}^\mathrm{TB}=1.08\,\mathrm{eV}$, respectively.
%This is not surprising since TB is known to also underestimate the quasiparticle band gaps of pristine AGNRs~\cite{son_energy_2006}.
%A similar effect was observed for pristine AGNRs for which TB is known to underestimate the band gap due to missing correlation effects~\cite{son_energy_2006}.
In addition to the two bands in the bulk region an additional unphysical zero-energy mode appears in the TB solution.\\
%\footnote{It is interesting to note that the TB result has a large contribution at $E=0$ in the bulk which is caused by the zero-energy end state that TB fails to damp correctly.} 
%Another important observation can be made for the UHF approximation which is known to qualitatively describe edge magnetism in free-standing ZGNRs~\cite{magda_room-temperature_2014,van_der_lit_suppression_2013,wang_giant_2016}. 
Next, we include mean-field effects within the UHF approximation. This does not lead to a considerable energy renormalization for the bulk and end states, 
%of the 7-9-AGNR,
%as compared to TB since it also underestimates the level of electronic correlations.
%\footnote{For $U=1.5J$ there is a small shift. However, as UHF is a mean field approach, of course the correlation induced renormalization is not captured correctly.}
but to a splitting of all three topological states which are localized at the end of the heterostructure, cf. Fig.~\ref{fig:exp_comparison}(c). The behaviour of the two non-zero energy end peaks is particularly surprising. While the splitting of the zero-energy edge peaks in ZGNRs is well known~\cite{magda_room-temperature_2014} and is attributed to magnetic instabilities at the Fermi level~\cite{pisani_electronic_2007}, the splitting of states at non-zero energy cannot be understood in this picture. In the experimental data where the system is on top of a screening Au(111) surface this effect is not observed. However, metallic substrates are known to suppress the splitting of zero-energy states in finite length pristine AGNRs~\cite{van_der_lit_suppression_2013} compared to insulating substrates~\cite{wang_giant_2016}.
The UHF result indicates that a splitting of all three end peaks, as seen in Fig.~\ref{fig:exp_comparison}(c), will emerge in measurements for an insulated heterojunction.\\
%This observation is remarkable since the splitting of the zero-energy edge peaks in ZGNRs is attributed to magnetic instabilities at the Fermi level~\cite{pisani_electronic_2007}. However, this can not explain the splitting of the non-zero end peaks observed here.
%Taking correlations into account using $GW$ with $U=2.5J$
%In contrast, our GW simulations 
Finally, including quasiparticle corrections within the $GW$ approximation
results in a considerable correlation-induced renormalization of both the bulk and end states. The 
%increased 
observed gaps of $E_\mathrm{g,bulk}^{GW}=0.71\,\mathrm{eV}$ and $E_\mathrm{g,end}^{GW}=1.35\,\mathrm{eV}$ are in excellent agreement with the experimental findings. Additionally, $GW$ correctly reduces the unphysical zero-energy contribution in the bulk and prevents the splitting of the end states through selfconsistent screening
%captures the asymmetry of the spectrum with the center of the lower bulk band at $-0.31\,\mathrm{eV}$ and the upper band at $0.4\,\mathrm{eV}$.
\footnote{One should note that the extreme broadening of the upper bulk band seen in the experiment is not captured by our simulations. The origin of this broadening is unclear at present, however it is probably caused by the experimental setup, i.e. the substrate or the tip.}.
%\footnote{Additionally, $GW$ reduces the unphysical contribution at $E=0$ in the bulk spectrum. This damping of the zero-energy end state is important to understand the splitting in Fig~\ref{fig:size}.}
%It is important to note that

\begin{figure}[t]
\includegraphics[width=0.48\textwidth]{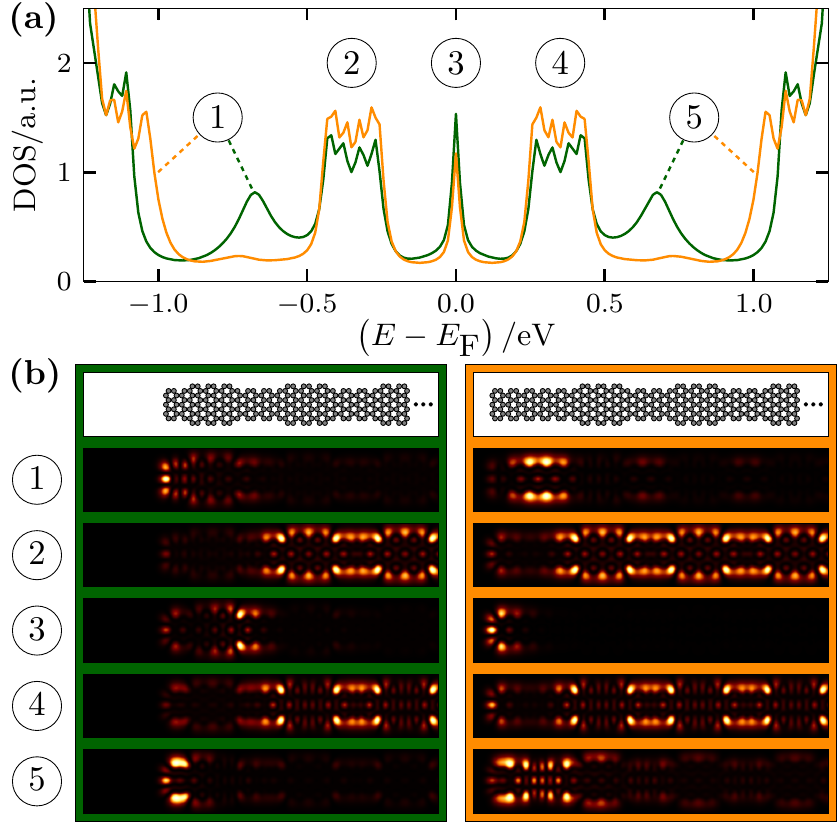}
\caption{\label{fig:extension} (a) Total DOS calculated with $GW$ and $U=2.5J$ for the two systems indicated in the top panel of (b). The green (orange) line corresponds to the left (right) system. The left system contains exactly six unit cells while the right system is extended additionally by ten zigzag lines on both sides. (b) d$I$/d$V$ maps for the same two systems. The maps labelled 1-5 correspond to the labelled peaks in (a). Only a small section of the ribbon is shown indicated by the three black dots in the top panel.}
\end{figure}

\textbf{Local correlations}.
%To reach a better understanding of the underlying mechanisms that define the above mentioned properties of the topological states, 
To understand the mechanisms causing the above mentioned renormalization and splitting of the topological states, 
%in the following, 
we next consider local correlations and magnetic polarizations. In Fig.~\ref{fig:localmoment}(b) we compare the local moment at site $i$
\begin{equation}
  \left\langle \chat{m}_i^2 \right\rangle = \left\langle \left( \chat{n}_{\uparrow,i} - \chat{n}_{\downarrow,i} \right)^2 \right\rangle = \rho_i - 2D_i\,,
\label{eq:moment}
\end{equation}
from UHF- and $GW$-simulations for the same system as in Fig.~\ref{fig:exp_comparison}. The local moment quantifies the local interaction energy and is a measure of the local magnetic polarization of the system. It is directly related to the local density $\rho_i$ and double occupation $D_i$ at site $i$ [cf. Eq.~(\ref{eq:moment})] which are strongly affected by electronic correlations.
%\footnote{The local moment for $GW$ is in general higher than for UHF (0.65 and 0.5, respectively) because UHF does not include correlations.}
Consequently, the local moment is, in general, higher for $GW$ than for UHF. In the latter case the local moment is peaked exactly at the sites where the zero-energy end state is localized, which can be confirmed by comparing to the LDOS in Fig.~\ref{fig:localmoment}(a). Considering the splitting of the zero-energy end peak in Fig.~\ref{fig:exp_comparison}(c) this is in agreement with previous mean-field calculations for ZGNRs \cite{pisani_electronic_2007,feldner_dynamical_2011} 
%where mean field theory predicts an antiferromagnetic ordering at opposing zigzag edges due to an instability of the zero-energy edge state.
where the magnetic instability of the zero-energy edge state gives rise to an antiferromagnetic ordering at opposing zigzag edges.
However, such an instability does not occur for the non-zero end states for which the local distribution only partially coincides with the local moment, cf. Fig~\ref{fig:localmoment}(a). Instead, the splitting of these states observed in Fig.~\ref{fig:exp_comparison}(c) originates from their hybridization with the zero-energy zigzag state which is further investigated in the discussion of Fig.~\ref{fig:extension}.\\
%However, here this same effect does not only result in a  splitting of the zero-energy peak but also of the two other non-zero energy peaks that are also localized in the end region of the ribbon, cf. Fig.~\ref{fig:exp_comparison}c.{\color{red}{maybe some more words on why the non-zero peaks split.}}
Strikingly, for $GW$ the local moment is increased on all edges of the heterostructure, which coincides with the regions where the topological states are localized, cf. Fig.~\ref{fig:extension}(b).
%with the highest peaks located at the zigzag edge and the heterojunctions between the 7- and 9-AGNR segments. {\bf In my print all the red blobs have essentially the same color, so I cannot confirm this statement.}
Consequently, the renormalization of the topological bulk and end peaks can be attributed to strongly localized correlations at the edges of the heterostructure.
%In fact, the shift of the non-zero end peaks is mainly influenced by the local interaction on the two zigzag edge sites marked in Fig.~\ref{fig:localmoment} (todo). Switching off the on-site interaction on these sides completely removes the shift seen in Fig~\ref{fig:exp_comparison}c (see supplement, todo).
%Another surprising observation
The topological states that extend across the boundary of the heterostructure result in increased magnetic polarization even at the armchair edges of the ribbon. This surprising finding is in contrast to the prediction of the UHF simulation in this work and other mean-field theory results~\cite{pisani_electronic_2007,fernandez-rossier_magnetism_2007,feldner_magnetism_2010,yazyev_theory_2011,golor_magnetic_2013,ortiz_electrical_2018} where typically considerable magnetic polarization is only observed at the zigzag edges of GNRs.

\begin{figure}[t]
\includegraphics[width=0.48\textwidth]{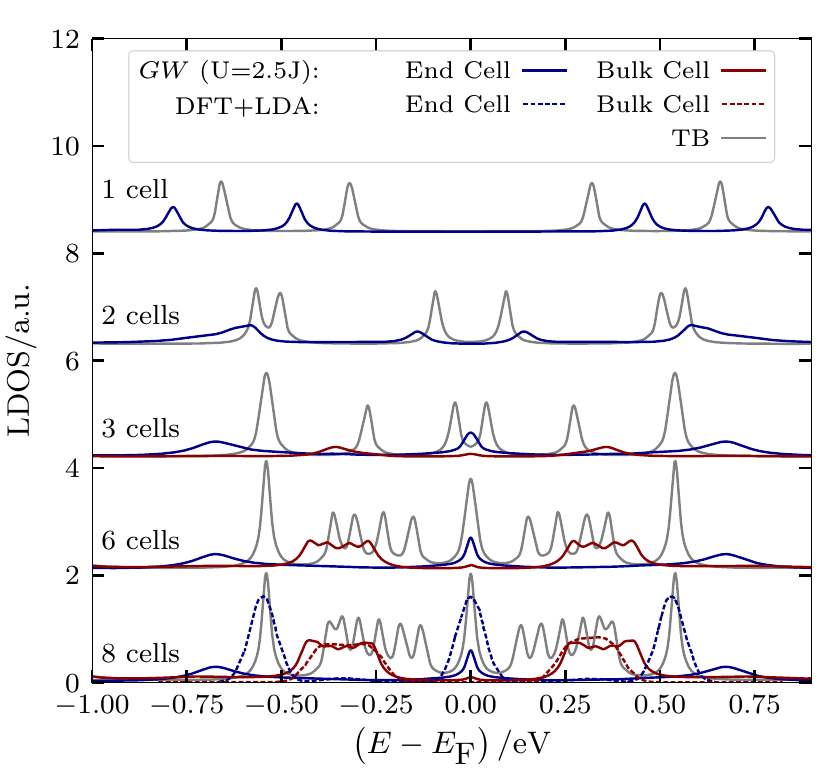}
\caption{\label{fig:size} LDOS for 7-9-AGNR heterostructures consisting of one to eight unit cells. The case of six unit cells is the one shown in Fig.~\ref{fig:exp_comparison}(a).  Red (blue) lines: %correspond to the LDOS in a 
bulk (end) cell results 
%of the ribbon 
as shown in Fig.~\ref{fig:exp_comparison}(a). Solid (dashed): $GW$ (LDA-DFT) calculations. The DFT results are taken from Ref.~\citenum{rizzo_topological_2018}. Solid grey lines: total DOS from TB calculations. The lines for different systems are shifted vertically for better comparison.}
\end{figure}
\textbf{Finite size effects}.
The GNR heterostructure discussed in Figs.~\ref{fig:exp_comparison} and~\ref{fig:localmoment} contains six unit cells, 
exactly as in the experiments. 
%that were performed for a system of similar size. 
In the following we explore the effect of the system size on the topological states.
%different system sizes will be investigated. 
In Fig.~\ref{fig:size} the local density of states (LDOS) of heterostructures containing one to eight unit cells is shown for the bulk and end cells comparing TB and $GW$ results. For the eight unit cell system, additionally, the LDA-DFT result of Ref.~\citenum{rizzo_topological_2018} is plotted to allow for a direct comparison to our results. The observed effects differ for the two regions of the system. The spectral weight of the bulk bands increases with system size. In fact, the TB results, that show the total DOS, indicate that the number of peaks in the bulk bands corresponds to the number of bulk unit cells. This is in agreement with the idea that the bulk bands form by hybridization of heterojunction states of adjacent bulk cells. For $GW$ the energies of the bulk states are renormalized and broadened by electronic correlations, as shown in Fig.~\ref{fig:exp_comparison}.\\

In the end cell the three topological states are stable for systems of three or more unit cells. For these large systems the states on both ends of the ribbon are separated by bulk cells. However, for smaller systems the states of opposing termini overlap and result in an additional splitting of the end states. In addition to this topological effect, the energies of the end states are also renormalized due to electronic correlations, in general resulting in higher energies of the states. However, interestingly, for the intermediate system of three unit cells the splitting of the zero-energy state is reduced for the correlated $GW$-result as compared to TB. This is the result of a competition of both aforementioned effects. Within $GW$ the spatial extension of the zero-energy state is strongly reduced, cf. Fig~\ref{fig:exp_comparison}(b) and SI, resulting in a significantly smaller overlap and finite-size splitting of the states on both ends of the system.\\

Comparing the results of $GW$ and LDA-DFT for the eight unit cell system reveals that LDA-DFT captures reasonably well the renormalization of the bulk band energies whereas it completely fails to describe the shift of the end states. This indicates that a correct characterization of these topological end states is particularly challenging and requires an accurate description of the underlying electronic correlations.
%For the eight unit cell system DFT+LDA shows good agreement with $GW$ for the bulk bands, while it completely misses the renormalization of the end peaks. \footnote{The strongly localized correlations at the end of the system are apparently not captured by DFT+LDA.} Thus, for a quantitative description of the end peaks quasiparticle corrections beyond DFT+LDA are mandatory.

%\textcolor{purple}{
%In contrast, LDA+DFT captures reasonably the renormalization of the bulk band energies whereas it completely fails to describe the shift of the end states
%{\bf shouldn't one have a reference here?}{\color{red}{The above statement cannot be verified by looking at the Rizzo paper since they never compare LDA and experiment directly. Thus, it's hard to find something to city here. This is why we have the second part of the sentence. }}
%, which can be confirmed by comparing the DOS results for the eight unit cell system in Fig.~\ref{fig:size}. This indicates that the correct characterization of these topological end states is especially challenging and requires a precise description of the underlying electronic correlations.
%}

\textbf{Topological end states}.
Since the topological states at the end of the heterostructure are found to be particularly sensitive to electron-electron interactions it is important to examine in detail the origin of these states.
The emergence of topological end states at the termini of GNR heterostructures can be explained by the $\mathbb{Z}_2$ invariant and the bulk-boundary correspondence of topological insulator theory~\cite{cao_topological_2017}. However, to get a better understanding of their %astonishing 
%remarkable 
properties, in the following, the specific mechanism that leads to the existence of these multiple end states will be analyzed. For this, in Fig.~\ref{fig:extension} the total DOS (a) and the d$I$/d$V$ maps (b) of the heterostructure containing six unit cells (green) are compared to the same system with an additional ten zigzag lines on each side of the ribbon (orange) for $GW$. The two systems are depicted in the topmost panel of Fig.~\ref{fig:extension}(b). Comparing the DOS it stands out that the high-energy end peaks, that are present in the six unit cell system, are strongly suppressed in the longer system. Instead a new high-energy peak emerges around $\pm1\,$eV. Additionally, the spectral weight of the bulk bands is increased for the longer system.
All these observations can be understood 
%by looking at 
from the d$I$/d$V$ maps of the states in Fig.~\ref{fig:extension}(b). For the longer system the zigzag edge at the end of the system is separated from the heterojunction of the first unit cell by a long 7-AGNR section. Due to this clear separation the three end states (1, 3, 5) resemble states observed in pristine 7-AGNR, see SI. Furthermore, the heterojunction states of all six unit cells hybridize to form the bulk bands (2, 4). In contrast, for the shorter system the adjacent states localized at the zigzag edge and at the heterojunctions of the first unit cell can overlap leading to three hybridized states in the termini of the system. As a consequence the bulk bands are localized only in the four inner bulk unit cells and do not occupy the end cells resulting in a reduced spectral weight in the DOS.\\

Splitting of peaks induced by the spatial overlap of the associated states typically appears in small finite systems, cf. Fig.~\ref{fig:size} for one unit cell. Strikingly, due to the close proximity of finite substructures, i.e. heterojunctions and ribbon edges, the same effect emerges at the termini of large heterostructures containing hundreds to thousands of atoms. Consequently, this observation is not restricted to the present 7-9-AGNR heterostructures. Instead, we predict that similar strongly correlated states exist also in an entire class of systems which meet the criteria for hosting hybridized end states, i.e. possess topological states close to localized edge states. The existence of such states is determined by the topology of the system~\cite{cao_topological_2017,lee_topological_2018,lin_topological_2018}.

{\bf Summary and discussion}.
We analyzed the influence of electronic correlations on the topological states of 7-9-AGNR heterostructures on Au(111). While the general topological structure of the system, previously predicted on the TB level~\cite{cao_topological_2017}, remains stable, additional new effects connected to the topological states emerge. Our $GW$ simulations reveal that strong local electronic correlations are present in both the edges of the bulk and the end region of the heterostructure resulting in increased magnetic moments in the zigzag and armchair edges. Strikingly, the spatially confined topological states of the termini are more severely affected by these correlations than the extended topological bulk bands. For the latter we found, by comparison to our $GW$ results, that LDA-DFT is able to reproduce the experimental d$I$/d$V$ measurements since quasiparticle corrections are weak, in this case. In contrast, the topological end states, emerging due to the hybridization of zigzag-edge and heterojunction states, are strongly renormalized and screened due to electronic correlations which gives rise to a large discrepancy between LDA-DFT and experimental energies. For free-standing heterostructures we find a new mechanism that leads to the splitting of non-zero energy states due to hybridization with a zero-energy state. These findings are not restricted to the specific system considered here but instead are expected to be present in similar GNR heterostructures that exhibit strongly localized topological states.

\begin{acknowledgement}

We acknowledge helpful discussions with Niclas Schl\"unzen. 
APJ is supported by the Danish National Research Foundation, Project DNRF103.
\end{acknowledgement}

%%%%%%%%%%%%%%%%%%%%%%%%%%%%%%%%%%%%%%%%%%%%%%%%%%%%%%%%%%%%%%%%%%%%%
%% The same is true for Supporting Information, which should use the
%% suppinfo environment.
%%%%%%%%%%%%%%%%%%%%%%%%%%%%%%%%%%%%%%%%%%%%%%%%%%%%%%%%%%%%%%%%%%%%%
\begin{suppinfo}

The following files are available free of charge.
\begin{itemize}
  \item suppinfo.pdf: Details on the Green function approach, the fitting of the Hubbard interaction $U$, damping of the zero-energy state, and a comparison to 7-AGNR states.
\end{itemize}

\end{suppinfo}

%%%%%%%%%%%%%%%%%%%%%%%%%%%%%%%%%%%%%%%%%%%%%%%%%%%%%%%%%%%%%%%%%%%%%
%% The appropriate \bibliography command should be placed here.
%% Notice that the class file automatically sets \bibliographystyle
%% and also names the section correctly.
%%%%%%%%%%%%%%%%%%%%%%%%%%%%%%%%%%%%%%%%%%%%%%%%%%%%%%%%%%%%%%%%%%%%%
\bibliography{library}

\end{document}